# Magneto-transport studies on $(Pr_{1/3}Sm_{2/3})_{2/3}A_{1/3}MnO_3$ (A = Ca, Sr and Ba) compounds

**Saket Asthana[1], D. Bahadur[1], A.K. Nigam[2] and S.K. Malik[2]**

[1]Department of Metallurgical Engineering and Materials Science, Indian Institute of Technology, IIT Bombay, Powai, Mumbai 400076, India

[2]Tata Institute of Fundamental Research, Colaba, Mumbai 400005, India

## Abstract

Magnetic and transport properties of $(Pr_{1/3}Sm_{2/3})_{2/3}A_{1/3}MnO_3$ (A = Ca, Sr and Ba) compounds, prepared by the citrate gel route, have been investigated. These compounds are found to crystallize in the orthorhombic structure. Charge ordering transport behavior is indicated only in Ca-substituted compound. The Sr- and Ba-substituted compounds show metal-insulator transition and semiconducting-like behavior, respectively. The magnetoresistance is highest in the Ba substituted compound. All the three samples show irreversibility in magnetization as a function of temperature in zero-field cooled (ZFC) and field cooled (FC) plots. The non-saturating magnetization, even at 5K and 4 Tesla field, are observed in Ca as well Ba-substituted compounds.

## Introduction

The mixed valent perovskite manganites of the general formula $Ln_{1-x}A_xMnO_3$ (Ln = rare earth and A = divalent alkaline earth cation) have created a great deal of interest because of their colossal magnetoresistance (CMR) behaviour. These systems also have technological importance such as in sensor applications, and especially for increasing data storage by increasing the sensitivity of hard disk drive read heads [1,2]. These manganites become ferromagnetic (FM) at an optimal value of x (or $Mn^{4+}$ content) and undergo metal-insulator (MI) transition around the ferromagnetic transition temperature. Effects of divalent alkaline-earth element substitution in the stoichiometric perovskite manganites $Ln_{1-x}A_xMn_{1-y}M_yO_3$ have been extensively studied [3,5]. These studies show that the Curie temperature, $T_C$, and the magnetoresistance are optimized for $Mn^{4+}$ content of about 33%. These properties are attributed to the double exchange (DE) interaction associated with electron hopping from $Mn^{3+}$ to $Mn^{4+}$. The double exchange interaction, which favors itinerant electron behaviour, is opposed by the Jahn-Teller (JT) distortion of the $Mn^{3+}$. Recent studies have shown that DE alone cannot explain the observed behavior in manganites [6] and other effects also play an important role. These include charge ordering, average A-site cationic radius $<r_A>$ [7,8], A-size cationic size mismatch [9,10], oxygen deficiency [11], electron-lattice coupling [12], polaron effect due to strong electron–phonon interaction arising from the Jahn-Teller distortion [6], etc. The average size of the A-site cation of these perovskites and the size mismatch at the A-site modify the Mn-O-Mn bond angle and affect the $e_g$ electron hopping between $Mn^{3+}$ and $Mn^{4+}$ degenerate states. The effect of ionic size variation can also be understood by the tolerance factor defined as, $t = (<r_A> + r_O)/\sqrt{2}\,(r_B+r_O)$, where $r_O$ and $r_B$ are radii of the oxygen and the B-site transition metal ions, respectively.

The electronic properties of the manganites can be tuned either by substituting cations at the A- or B-sites or by varying the oxygen content in the regular perovskites structure [13-15]. The A-site cation is responsible for structural distortions while the B-site cation is responsible for the magnetic interactions. Mixing cations of different charges at the A-site is the most straightforward experimental method for systematically tuning the properties of these materials. There are many ways to achieve this with $Ln^{3+}/M^{2+}$ combinations (Ln = rare earth and M = Ca, Ba Sr and Pb) or with rare earth ion combinations. This has led to efforts to unify the properties of compositions from different chemical phase diagrams within a single framework by using a simple ionic description of the A-site cations [16].

A lot of work has been done on manganites with the single rare earth ion at the A-site. In the present work, we have undertaken to substitute two rare earth ions, namely, Pr and Sm at the A-site in a fixed ratio and vary the alkaline earth ion. In this context, we have chosen the series $(Pr_{1/3}Sm_{2/3})_{2/3}A_{1/3}MnO_3$ (A = Ca, Sr and Ba) for present studies.

## Experimental Details

The polycrystalline $(Pr_{1/3}Sm_{2/3})_{2/3}A_{1/3}MnO_3$ (A = Ca, Sr and Ba) samples were synthesized by the chemical citrate-gel route using high purity $Pr_6O_{11}$, $Sm_2O_3$, $CaCO_3$, $BaCO_3$, $SrCO_3$ and Mn-acetate. The as prepared powders were calcined at 1000°C in air for two hours. The powders were pelletized in the form of rectangular bars and sintered at 1200°C in air. X-ray diffraction patterns of the samples were recorded using Cu-$K_\alpha$ radiation (PW 3040/60 Philips, PANalytical). Resistivity measurements at different applied magnetic fields were carried out from 320 to 5 K using the standard four-probe dc method (PPMS, Quantum Design). Magnetic measurements were made using vibrating sample magnetometer (Oxford, Maglab VSM) at different fields and in the temperature range of 300 K to 5 K.

## Results and Discussion

### Crystal Structure

Figure 1 shows the X-ray diffraction patterns for $(Pr_{1/3}Sm_{2/3})_{2/3}A_{1/3}MnO_3$ (A = Ca, Sr and Ba) compositions. All the lines in the patterns could be indexed on the basis of an orthorhombic structure (space group Pnma No. 62). In Ca-substituted sample, splitting of some of the X-ray lines is observed which is due to increased orthorhombicity of this compound (compared to that of other two). This in turn is due to smaller ionic size of Ca ion (1.18 Å) in comparison to those of Sr(1.31 Å) and Ba(1.47Å) ions. The average A-site ionic radius, $< r_A >$ varies from 1.158 Å for Ca-, 1.202 Å for Sr- and 1.255 Å for Ba-substituted samples. The mean radius has been calculated using the coordination number nine [17]. The tolerance factors for Ca, Sr and Ba containing compounds are 0.9018, 0.9173 and 0.9360, respectively. This indicates that the structure tends towards pseudo-cubic symmetry as one goes from Ca to Ba substituent. The cell volume decreases with decreasing A-site ionic radius.

The cell and structural parameters of all the $(Pr_{1/3}Sm_{2/3})_{2/3}A_{1/3}MnO_3$ (A = Ca, Sr and Ba) compounds were refined by the Rietveld method using the computer code FULLPROF [18]. The refinement was carried out in the space group Pnma (No. 62) with the following atomic positions: Pr/Sm/A: 4c(x,y,1/4), Mn: 4b(0,0,1/2), O(1): 4c(x,1/4,z) and O(2):(x,y,z). The refined lattice parameters are given in Table 1. The oxygen positions, derived from the refinement have high errors due to the presence of strong scatterers Pr and Sm ions [19]. The best-fit $\chi^2$ values for Ca, Sr and Ba-substituted compounds are 2.89, 1.54 and 2.62, respectively. A typical refined pattern for one sample, namely, the Sr-substituted compound, is shown in Fig. 2.

### Electrical Resistivity

The temperature dependence of electrical resistivity, ρ(T), for all the $(Pr_{1/3}Sm_{2/3})_{2/3}A_{1/3}MnO_3$ (A = Ca, Sr and Ba) compounds is shown in Fig. 3. The resistivity values are very high for Ca and Ba- substituted compounds over the entire temperature range from 320 K to 50 K and these compounds show semiconducting like behaviour (d$\rho$/dT<1). The typical metal-insulator transition is observed only for Sr-substituted compound.

In Ca-substituted sample, $(Pr_{1/3}Sm_{2/3})_{2/3}Ca_{1/3}MnO_3$, a small anomaly in the dln$\rho$/d$(k_BT)^{-1}$ plot is seen around 195 K (inset of Fig. 3) which is a typical feature of charge ordering as also reported for the parent compound $Pr_{0.67}Ca_{0.33}MnO_3$ [20,21]. It has been reported that the compounds having tolerance factor around or less than 0.907, may exhibit charge order phenomena [22]. The tolerance factor for Ca-substituted sample is 0.9018 and therefore it is probably the reason for its charge ordering behavior. As mentioned earlier, the size of the $Ca^{2+}$ ion is smaller than those of the $Ba^{2+}$ and $Sr^{2+}$ ions, which in turn creates more structural distortions in the orthorhombic lattice of the Ca-substituted compound. Because of this, the Mn-O-Mn bond angle deviates more from $180^0$ in this compound. The transfer integral (t) of the $e_g$ electrons hopping from $Mn^{3+}$ to $Mn^{4+}$, defined as $t = t_0 \cos(\theta/2)$ (where $t_0$ is the normal transfer integral and $\theta$ is the angle between two neighbouring spin directions [23]) reduces more drastically in the case of Ca- and Ba-substituted compounds as compared to that in Sr- substituted compound. This leads to suppression of the DE interaction between the $Mn^{3+}$ and $Mn^{4+}$ and to higher resistivity values in Ca- and Ba- substituted compounds. However, anomaly in dln$\rho$/d$(k_BT)^{-1}$ is absent in Ba-substituted compound.

The Sr- substituted sample, $(Pr_{1/3}Sm_{2/3})_{2/3}Sr_{1/3}MnO_3$, exhibits the M-I transition similar to that observed in parent $Pr_{0.7}Sr_{0.3}MnO_3$ (PSMO) [24], but the $T_{MI}$ is shifted towards lower temperatures. The $T_{MI}$ for Sr-substituted sample, estimated from the average value between minima and maxima of d$\rho$/dT plot, is 156 K which is in between that of 265 K for the parent compounds $Pr_{0.7}Sr_{0.3}MnO_3$ and 100K for $Sm_{2/3}Sr_{1/3}MnO_3$ (SSMO) [25]**.** The drop in $T_{MI}$ is due to the variation in the A-site ionic radius $\langle r_A \rangle$ of the compounds. The $\langle r_A \rangle$ for the Sr- substituted sample, $(Pr_{1/3}Sm_{2/3})_{2/3}Sr_{1/3}MnO_3$, is found to be 1.202 Å which is in between the $\langle r_A \rangle$ of PSMO (1.223 Å) and SSMO (1.191 Å). The difference between the A-site ionic radius of $(Pr_{1/3}Sm_{2/3})_{2/3}Sr_{1/3}MnO_3$ and SSMO is lower (0.011 Å) as compared to that between $(Pr_{1/3}Sm_{2/3})_{2/3}Sr_{1/3}MnO_3$ and PSMO (0.021 Å). Therefore, the $T_{MI}$ of 156K for $(Pr_{1/3}Sm_{2/3})_{2/3}Sr_{1/3}MnO_3$ sample is closer to $T_{MI}$ of 100K for SSMO rather than to $T_{MI}$ of 265K for PSMO.

Although metal to insulator transition is observed only in Sr-substitued $(Pr_{1/3}Sm_{2/3})_{2/3}Sr_{1/3}MnO_3$ compound, all the $(Pr_{1/3}Sm_{2/3})_{2/3}A_{1/3}MnO_3$ (A = Ca, Sr and Ba) compounds show paramagnetic (PM) to ferromagnetic (FM) transition (see in magnetization section). In Sr-substituted $(Pr_{1/3}Sm_{2/3})_{2/3}Sr_{1/3}MnO_3$ compound, the $T_{MI}$ and $T_C$ are very close to each other, which is evidence of long range ferromagnetism. Thus for Ba and Ca –substituted compounds (where MI is not observed), long range ferromagnetism, due to double exchange, disappears and consequently hinders metallicity leading to monotonous increase in resistivity [26].

The transport data in semiconducting regime of $(Pr_{1/3}Sm_{2/3})_{2/3}A_{1/3}MnO_3$ (A = Ca, Sr and Ba) compounds has been analyzed by Mott Variable Range Hopping model (Mott-VRH) [27] according to which

$$\rho = \rho_{0m} \exp (T_{0m}/T)^{1/4} \qquad (1)$$

Here $\rho_{0m}$ is the Mott residual resistivity and $T_{0m}$ is the Mott characteristic temperature. Equation (1) is found to fit the data well as shown in Fig. 3b. The values of $\rho_{0m}$ and $T_{0m}$ are listed in Table 2. A small deviation in theoretical fit is observed in Ca-substituted system at the charge ordering temperature ($T_{CO}$ = 195 K) as indicated by arrow in Fig. 3b.

The metallic region in Sr-substituted $(Pr_{1/3}Sm_{2/3})_{2/3}Sr_{1/3}MnO_3$ system has been fitted to a Zener-double exchange polynomial

$$\rho = \rho_0 + \rho_2 T^2 + \rho_n T^n \qquad (2)$$

where, $\rho_0$ is the temperature independent residual resistivity due to scattering by impurities, defects, grain boundaries and domain walls. The second term with coefficient $\rho_2$ is ascribed to the electron-electron and electron-phonon scattering mechanism [28] and the third term with coefficient $\rho_n$ corresponds to the electron-magnon scattering [27]. The value of n is found to be 7.2 in the Sr- substituted system. The fitted plot is shown in the inset of Fig. 3b. The typical fitted parameters are summarized in Table 2.

**Magneto-Resistance**

Figure 4a shows the electrical resistivity behaviour of Sr- substituted, $(Pr_{1/3}Sm_{2/3})_{2/3}Sr_{1/3}MnO_3$ sample in different applied magnetic fields. In the presence of an external field, the resistivity decreases significantly and $T_{MI}$ shifts to a higher value from 156 K in zero field to 192 K in 8 T field. This suggests that the external magnetic field facilitates the hopping of $e_g$ electron between the neighbouring Mn ions, which agrees with the DE model [29]. The magnetoresistance (MR) is defined as

$$\% \text{ MR} = 100 \times [\rho(H, T)- \rho(0,T)]/ [\rho(0, T)] \qquad (3)$$

where, ρ(H,T) and ρ(0,T) are the resistivities at a temperature T, in applied magnetic field H and in zero applied magnetic field, respectively. The temperature dependence of MR for Sr-substituted compound is showing in Fig. 4b. The MR is quite high (> 95 % at 8T) at the metal insulator transition temperature. The manganese ions are ferromagnetically ordered below $T_C$, therefore, within a single magnetic domain, the $e_g$ electron transfer between $Mn^{3+}$ and $Mn^{4+}$ ions is easy. The pairs of $Mn^{3+}$ and $Mn^{4+}$ spins, which may not be parallel in the vicinity of domain wall boundaries, will act as a hindrance for electron transport. The magnetic domains tend to align along the field direction in the presence of sufficiently strong magnetic field. As a result, hopping of electrons become easy across the domain wall boundaries and resistivity decreases, which in turn leads to significant MR at low temperatures. Unlike in $La_{2/3}Sr_{1/3}MnO_3$, the MR in Sr-substituted compound does not drop to a low value at low temperatures but rather remains high and weakly temperature dependent below 100 K due to the inter grain tunneling at low temperatures [15].

The temperature dependence of MR for Ba- and Ca- substituted $(Pr_{1/3}Sm_{2/3})_{2/3}A_{1/3}MnO_3$ compounds is shown in Fig. 4c. It is remarkable to note that for Ba substituted sample, MR is ~99% at low temperatures in 8 Tesla magnetic fields while the same is around 60% for Ca-substituted sample though the temperature dependence of MR exhibits similar nature in the two. The continuous rise in MR with decreasing temperature and its large value in Ba- and Ca- substituted samples appear to be due to electronic phase separation, where presumably the ferromagnetic (FM) clusters start growing at the expense of anti ferromagnetic clusters with decreasing temperature in high field of 8T. Similar MR behavior has been observed in $(La_{0.6}Ho_{0.4})_{0.7}Sr_{0.3}MnO_3$ [30].

**Magnetization**

The plots of magnetization (M) as a function of temperature (T) for $(Pr_{1/3}Sm_{2/3})_{2/3}A_{1/3}MnO_3$ (A = Ca, Sr and Ba) compounds are shown in Fig.5 a-c. In Ca-substituted sample, at low fields (0.01T), the magnetization (M) shows irreversibility between the zero-field cooled (ZFC) and the field cooled (FC) state below 150 K. This irreversibility becomes very strong below around 40 K. At higher field (above 0.2 T), as shown in the inset (a) of Fig.5a, the irreversibility is observed only below 40 K. The magnetic transition temperatures have been determined from the minima in the dM/dT plots. In $Pr_{0.65}Ca_{0.35}MnO_3$, a charge ordering (CO) transition at 230 followed by AF transition around 165 K has been reported [31]. In the Ca- substituted sample, $(Pr_{1/3}Sm_{2/3})_{2/3}Ca_{1/3}MnO_3$, we observe a weak signature of AF transition below 150 K, where also setting in of irreversibility between ZFC and FC magnetization is seen. The appearance of irreversibility in M could be attributed to magnetic frustration generated due to competing FM and AFM moments as found in the neutron diffraction studies on $Pr_{0.7}Ca_{0.3}MnO_3$ [32].

For Ba-substituted $(Pr_{1/3}Sm_{2/3})_{2/3}Ba_{1/3}MnO_3$ sample, the irreversibility between ZFC and FC magnetization is seen below 50 K both in 0.01T and 0.1T field (Fig. 5c). This behavior is similar to that found in cluster glasses. In the case of Sr-substituted sample, a ferromagnetic transition is observed at $T_c$=158 K ( Fig. 5b), which is same as the metal-insulator transition temperature in zero field. Below $T_C$, irreversibility between ZFC and FC magnetization is observed, which could be due to domain wall pinning. Further, below around 25 K, M(T) drops presumably due to competing superexchange and double exchange interactions resulting in the formation of localized spin clusters.

Figure 6 shows the isothermal magnetization at 5 K for $(Pr_{1/3}Sm_{2/3})_{2/3}A_{1/3}MnO_3$ (A = Ca , Sr and Ba) compounds. The magnetization of Sr-substituted compound shows near saturation and its value is 3.62 $\mu_B$/Mn , which is close to the theoretical value (3.7 $\mu_B$/Mn). The magnetization of Ba and Ca-substituted samples do not saturate even in a field of 4 T. The Ba-substituted sample shows a large remanence in M(H=0) indicating presence of strong FM phase in it. The non-saturating near linear behavior is observed in Ca-substituted system at 5 K even up to 5 T magnetic field, which is expected in a AF or spin-glass like state. The non saturating M-H behavior for Ca and Ba–substituted samples may also result due to the presence of well-separated charge delocalized ferromagnetic clusters in the antiferromagnetic insulating matrix [33]. These clusters start growing within the matrix with increasing magnetic field.

The field dependence of magnetization as shown in Fig. 6 supports the MR behavior, i.e Sr-substituted sample is ferromagnetic, the Ba-substituted sample has a spin glass (or cluster glass) type of disorder giving rise to a non saturating behavior even at 8 T (Presumably in this compound electronic phase separation is quite significant). In the Ca-substituted compound, the charge ordered state is much more robust, having a nearly AF state. However the large MR (~ 60 %) at low temperature suggests that, there is electronic phase separation in this compound too.

## Conclusions

We have investigated the effect of alkaline earth ion substitution on the structural and magneto-transport properties of $(Pr_{1/3}Sm_{2/3})_{2/3}A_{1/3}MnO_3$ (A = Ca, Sr and Ba) compounds. Synthesized samples crystallize in the orthorhombic structure (space group Pnma No. 62). The Ba- and Ca-substituted, $(Pr_{1/3}Sm_{2/3})_{2/3}A_{1/3}MnO_3$, (A = Ca and Ba) samples show an insulating behavior. However, a small anomaly due to charge ordering is observed around 195 K for Ca- substituted sample only. Unlike Ba- and Ca-substituted samples, the Sr- substituted sample shows metal to insulator transition. Spin glass/cluster like behavior is observed in Ca- and Ba-substituted samples, which is also supported by the non saturating behavior in M-H plots. The non saturating MR value at low temperatures is observed in the case of Ca and Ba-substituted samples. In addition, Ba-substituted sample exhibits a very high MR (~99%) in 8T magnetic field. The magnetic behavior is attributed due to the electronic phase separation.

## Acknowledgement

Two of authors (SA and DB), are thankful to the Department of Science and Technology (DST) India for support of the project.

## Figure Captions

**Fig. 1**. X-ray diffraction patterns of the series $(Pr_{1/3}Sm_{2/3})_{2/3}A_{1/3}MnO_3$ (A=Ca, Sr and Ba) showing the single phase nature of the compounds.

**Fig. 2.** Rietveld refined, X-ray pattern of $(Pr_{1/3}Sm_{2/3})_{2/3}Sr_{1/3}MnO_3$ using the space group Pnma No.62 ($\chi^2 = 1.54$).

**Fig. 3a** Variation of resistivity with temperature for $(Pr_{1/3}Sm_{2/3})_{2/3}A_{1/3}MnO_3$ (A = Ca, Sr and Ba). The $d\ln\rho/d(k_BT)^{-1}$ vs. T plot for Ca-substituted $(Pr_{1/3}Sm_{2/3})_{2/3}Ca_{1/3}MnO_3$ (inset) shows the anomaly at the charge order temperature (marked by an arrow).

**Fig. 3b** Theoretical fit of the transport data using Mott VRH model for $(Pr_{1/3}Sm_{2/3})_{2/3}A_{1/3}MnO_3$ (A = Ca, Sr and Ba). Charge Ordering point in Ca-substituted system is indicated by an arrow. The inset shows theoretical fit of the transport data for Sr-substituted system in metallic regime.

**Fig. 4a** Variation of resistivity with temperature at different fields for Sr-substituted $(Pr_{1/3}Sm_{2/3})_{2/3}Sr_{1/3}MnO_{3,}$ sample.

**Fig. 4b** Variation of negative MR% with temperature for Sr-substituted, $(Pr_{1/3}Sm_{2/3})_{2/3}Sr_{1/3}MnO_{3,}$ sample. The field dependence of MR at different temperatures is shown in the inset

**Fig. 4c** Variation of negative MR% with temperature for $(Pr_{1/3}Sm_{2/3})_{2/3}A_{1/3}MnO_3$ (A = Ba and Ca) samples.

**Fig. 5a** Temperature dependence of magnetization in a field of 0.01 T in ZFC and FC runs for Ca-substituted $(Pr_{1/3}Sm_{2/3})_{2/3}Ca_{1/3}MnO_{3,}$ sample. The inset shows the M(T) plot in a field of 0.2 T for the same.

**Fig. 5b** FC and ZFC magnetization plots for the Sr- substituted $(Pr_{1/3}Sm_{2/3})_{2/3}Sr_{1/3}MnO_3$ sample in a field of 0.01 T.

**Fig. 5c** Variation of magnetization with temperature for the Ba-substituted, $(Pr_{1/3}Sm_{2/3})_{2/3}Ba_{1/3}MnO_3$ sample in fields of 0.01 T and 0.1 T.

**Fig. 6** The isothermal variation of magnetization with field for the series $(Pr_{1/3}Sm_{2/3})_{2/3}A_{1/3}MnO_3$ ( A = Ca, Sr and Ba) at 5 K.

## Table Captions

**Table 1.** Refined lattice parameters using Rietveld method, tolerance factors, and $T_C$ for the series, $(Pr_{1/3}Sm_{2/3})_{2/3}A_{1/3}MnO_3$ ( A = Ca, Sr and Ba)

**Table 2.** Parameters obtained from experimental and theoretical analysis

# Table 1

| Systems | a (Å) | b (Å) | c (Å) | V (Å)$^3$ | t | $T_C$ (K) |
|---|---|---|---|---|---|---|
| $(Pr_{1/3}Sm_{2/3})_{2/3}$ $Ca_{1/3}MnO_3$ | 5.4780(7) | 7.6501(0) | 5.4067(8) | 226.58(5) | 0.9018 | 40 |
| $(Pr_{1/3}Sm_{2/3})_{2/3}$ $Sr_{1/3}MnO_3$ | 5.4489(5) | 7.6910(8) | 5.4525(6) | 228.50(7) | 0.9173 | 158 |
| $(Pr_{1/3}Sm_{2/3})_{2/3}$ $Ba_{1/3}MnO_3$ | 5.4949(1) | 7.7740(7) | 5.4937(8) | 234.68(3) | 0.9360 | 64 |

# Table 2

| Systems | $(Pr_{1/3}Sm_{2/3})_{2/3}Ca_{1/3}MnO_3$ | $(Pr_{1/3}Sm_{2/3})_{2/3}Sr_{1/3}MnO_3$ | $(Pr_{1/3}Sm_{2/3})_{2/3}Ba_{1/3}MnO_3$ |
|---|---|---|---|
| $T_{MI}/T_{CO}$ (K) | $T_{CO}$ = 195 | $T_{MI}$ = 156 K | - |
| $\rho_{300\,K}$ (Ω cm) | 5.42 | 0.17 | 0.32 |
| $\rho_0$ (Ω cm) | - | 0.28 | - |
| $\rho_2$( Ω cm $K^{-2}$) | - | 3 x $10^{-5}$ | - |
| $\rho_{7.2}$ (Ω cm $K^{-7.2}$) | - | 1.5 x $10^{-15}$ | - |
| $\rho_{0m}$ (Ω cm) | 1.60 | 0.05 | 0.11 |
| $T_{0m}$ (x $10^8$ K) | 1.72 | 1.62 | 1.57 |